\documentclass[twocolumn,amsmath,amssymb,a4paper,superscriptaddress]{revtex4}

\usepackage{graphicx}

\usepackage[mediumspace,mediumqspace,squaren,Gray]{SIunits}

\begin{document}

\title{Excitonic Fano Resonance in Freestanding Graphene}

\author{Dong-Hun Chae\footnote{These authors contributed equally.}}

\affiliation
{Max Planck Institute for Solid State Research, Heisenbergstrasse 1, D-70569 Stuttgart,
Germany}

\author{Tobias Utikal$^\star$}

\affiliation{4$^\mathit{th}$ Physics Institute and Research Center SCOPE, University of
Stuttgart, Pfaffenwaldring 57, D-70550 Stuttgart, Germany}

\affiliation
{Max Planck Institute for Solid State Research, Heisenbergstrasse 1, D-70569 Stuttgart,
Germany}

\author{Siegfried Weisenburger}

\affiliation{4$^\mathit{th}$ Physics Institute and Research Center SCOPE, University of
Stuttgart, Pfaffenwaldring 57, D-70550 Stuttgart, Germany}

\affiliation
{Max Planck Institute for Solid State Research, Heisenbergstrasse 1, D-70569 Stuttgart,
Germany}

\author{Harald Giessen}

\affiliation{4$^\mathit{th}$ Physics Institute and Research Center SCOPE, University of
Stuttgart, Pfaffenwaldring 57, D-70550 Stuttgart, Germany}

\author{Klaus v. Klitzing}

\affiliation
{Max Planck Institute for Solid State Research, Heisenbergstrasse 1, D-70569 Stuttgart,
Germany}

\author{Markus Lippitz}
\email{m.lippitz@physik.uni-stuttgart.de}

\affiliation{4$^\mathit{th}$ Physics Institute and Research Center SCOPE, University of
Stuttgart, Pfaffenwaldring 57, D-70550 Stuttgart, Germany}

\affiliation
{Max Planck Institute for Solid State Research, Heisenbergstrasse 1, D-70569 Stuttgart,
Germany}

\author{Jurgen Smet}
\email{j.smet@physik.uni-stuttgart.de}

\affiliation
{Max Planck Institute for Solid State Research, Heisenbergstrasse 1, D-70569 Stuttgart,
Germany}

\date{\today}


\begin{abstract}
We investigate the role of electron-hole correlations in the absorption of freestanding monolayer and bilayer graphene using optical transmission spectroscopy from 1.5 to 5.5 eV. Line shape analysis demonstrates that the ultraviolet region is dominated by an asymmetric Fano resonance. We attribute this to an excitonic resonance that forms near the van-Hove singularity at the saddle point of the band structure and couples to the Dirac continuum. The Fano model quantitatively describes the experimental data all the way down to the infrared. In contrast, the common non-interacting particle picture cannot describe our data. These results suggest a profound connection between the absorption properties and the topology of the graphene band structure.

\end{abstract}

\maketitle

\begin{figure}[tbp]
  \centering
  \includegraphics[width=0.7\columnwidth]{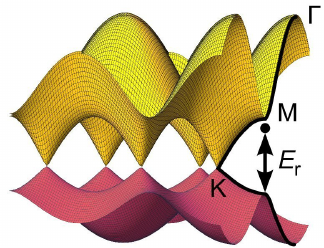}
  \caption{Electronic band structure of graphene with the saddle point (M) and an illustration of the excitonic state (dot).  }
    \label{fig:BZ}
\end{figure}

The material properties and the atomic structure of graphene are intimately connected. Most electronic effects can be understood by the unique band structure deduced from a tight-binding model of uncorrelated electrons [1]. A prominent example is the constant optical absorption for photon energies in the infrared wavelength range. It is a consequence of the linear dispersion relation near the K points in the Brillouin zone, the so-called Dirac cones. The absorption is given by fundamental constants alone as the product of the fine structure constant in vacuum  $\alpha \approx 1/137$ and  $\pi$  [2,3,4], and it is independent of the velocity of the Dirac fermions. Here, we demonstrate experimentally by line shape analysis that a single-particle model cannot describe the absorption spectrum of freestanding graphene in the visible and ultraviolet spectral region. The saddle point (M) in the band structure (see Fig.~\ref{fig:BZ}) causes a van-Hove singularity with a divergent density of states, allowing for a strong optical transition [5]. In this case, electron-hole correlations can lead to effects beyond the single-particle picture. An excitonic resonance at an energy slightly below the van-Hove singularity becomes possible. At a saddle point, the excitonic resonance takes a Fano shape as the discrete exciton state couples to the continuum formed by the band descending from the saddle point [5,6]. In the following we show that the Fano model of the excitonic resonance describes the complete optical spectrum of graphene from the ultraviolet all the way down to the infrared part of the electromagnetic spectrum.

A Fano resonance occurs when a discrete state couples to a continuum of states [6].
The resulting spectrum has the form
\begin{equation}
 A_{\rm{Fano}}(E) = C \; \left( 1 + \frac{q^2 - 1}{1 + s^2}   + \frac{2 q s}{1 +
s^2} \right)
 = C \; \frac{(s + q)^2}{1 + s^2} \label{eq:fano}
 \end{equation}
 where
 \begin{equation*}
 s = \frac{E - E_r}{\gamma / 2} \quad .
\end{equation*}
The damping rate of the resonance at an energy $E_r$ is quantified by the
line width $\gamma$. The asymmetric spectral shape is determined by
the unit-free Fano parameter $q$, which describes a ratio between the
transition probabilities to the discrete level and  a state in the continuum. $C$ is an overall scaling
factor. The three terms in the bracket may be interpreted as a transition into
the continuum, as a
Lorentzian associated with the discrete resonance, and as an
interference term, respectively [6]. 

\begin{figure*}
  \centering
  \includegraphics[width=2\columnwidth]{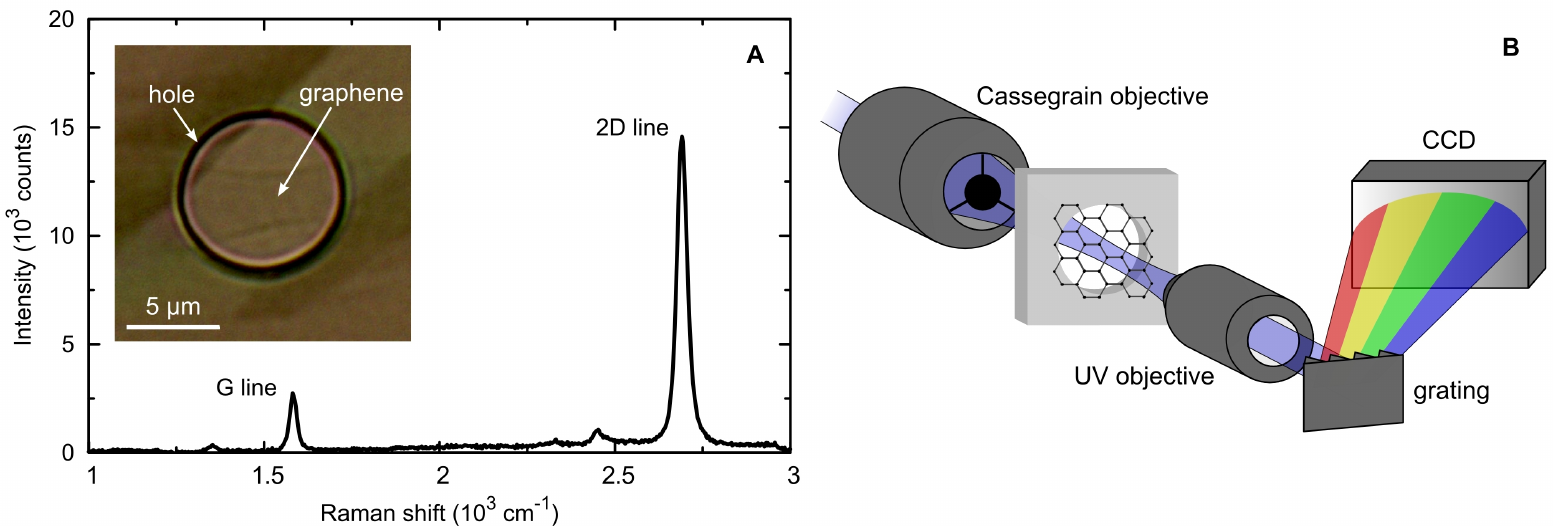}
  \caption{({\bf A}) Raman spectrum of a typical freestanding graphene sample.
Inset: Optical microscope image of a measured
graphene layer. This image was taken in transmission mode. ({\bf B}) Schematic
view of the experimental setup.}
  \label{fig:setup}
\end{figure*}

The single particle picture excludes electron-hole correlations. The
spectral shape of the interband transition near a saddle point is
proportional to the joint density of states. For a saddle point in a
two-dimensional system, the joint density of states is given by
[5] 
\begin{equation*}
 D(E) \propto - \ln( \left| E - E_r | \right) \quad .
\end{equation*}
To take the inhomogeneous broadening into account, we convolute the resulting
lineshape with a Gaussian function of variable width $\gamma$. The constant
absorption stemming from the Dirac cones is modeled as a constant offset $B$. The alternative single-particle model is thus given by
\begin{align}
& A_{\rm{single particle}}(E) = \label{eq:singleparticle} \\
  &C  \; \left(B + \exp\left[-
\frac{(E - E_r)^2 }{\gamma^2} \right] \otimes   \left[-\ln( \left| E
- E_r | \right)\right]\right) \quad ,  \notag
\end{align}
where $\otimes$ denotes a convolution. Note that both models have
the same number of free parameters.

Our samples are freestanding graphene monolayer and bilayer.
The  samples are fabricated by transferring graphene
from a silicon substrate to an aperture in a polymer resist (see Fig.~\ref{fig:setup}A
inset). The graphene flakes are first prepared on an oxidized silicon
substrate using mechanical exfoliation of natural graphite. Graphene
and bilayer graphene are identified by the optical contrast in a
microscope image [7] 
and by Raman spectroscopy [8]. 
The samples are spin-coated with a
$\unit{500}{\nano\metre}$ thick layer of poly(methyl methacrylate)
(PMMA). Disk shapes with a diameter of $\unit{8}{\micro\metre}$ are
written on top of the flakes using electron beam lithography. After
development, the PMMA layer with the $\unit{8}{\micro\metre}$
aperture together with the graphene flake are removed from the
substrate by etching the silicon dioxide in a 5~\% NaOH aqueous
solution at $\unit{90}{\degreecelsius}$.

We characterize our samples by Raman spectroscopy. Figure \ref{fig:setup}A shows a Raman spectrum of a typical freestanding graphene
sample. The symmetric 2D line at about
$\unit{2700}{\centi\reciprocal\metre}$ is the hallmark of a
monolayer [8]. 
The position of the G line at
approximately $\unit{1582}{\centi\reciprocal\metre}$ suggests a
negligible level of doping [9] 
and marginal strain [10] 
introduced during sample fabrication.


We measure the optical transmission of freestanding monolayer and bilayer graphene in a
confocal microscope. Figure.~\ref{fig:setup}B depicts a schematic of
the experimental setup. The light source combines a tungsten halogen
bulb and a deuterium lamp. We implement point illumination by using
the end-face of an optical quartz fiber as a confocal pinhole. The
light is then focused onto the sample by an all-reflective
Cassegrain objective (Davin Optronics, 74x) with a numerical
aperture (NA) of 0.65. A UV microscope objective (Zeiss Ultrafluar,
100x, NA 1.2) recollects the transmitted light which is then
directed to the spectrometer, consisting of a monochromator and a
liquid nitrogen-cooled CCD camera as detector. Note that we use the
Ultrafluar objective without glycerol immersion in contrast to its
specifications.

This assembly allows us to take spectra over a
photon energy range from $\unit{1.5}{\electronvolt}$ to
$\unit{5.5}{\electronvolt}$ with a spatial resolution on the sample
better than $\unit{1.5}{\micro\metre}$.
Due to the limited spectral bandwidth
of each spectrometer grating, the whole spectrum is measured by concatenating
data
from three exposures using two different gratings. The spectral resolution is
about
$\unit{10}{\milli\electronvolt}$.

The transmittance $T$ is determined by the ratio of the transmitted
light intensity through the graphene membrane to the transmitted
intensity when the layer is replaced by an empty reference aperture
nearby. The transmittance for bilayer graphene is measured in the
same way. The weak absorption of the mono-atomic film implies a
negligible reflectance, hence the absorbance $A$ can be
written as $A = 1 - T$.

\begin{figure*}
  \centering
  \includegraphics[width=1.5\columnwidth]{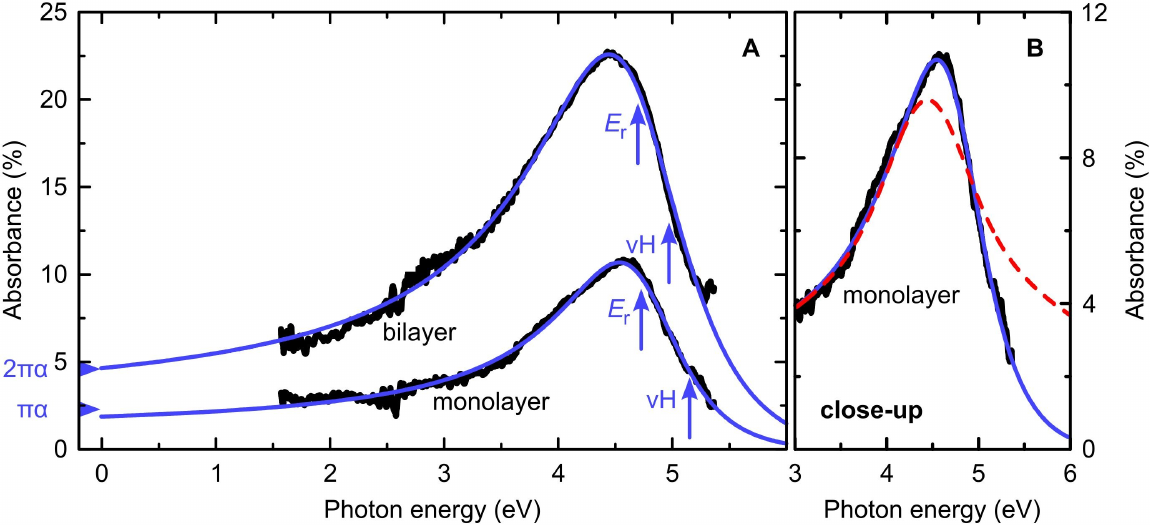}
  \caption{(A) Absorbance (= 1 - Transmittance) spectra of freestanding monolayer and bilayer graphene (black thick lines) are well described by a Fano model (blue thin lines). The difference between the resonance energy $E_r$ and the van-Hove singularity (vH) determines the exciton binding energy.  (B) Close-up of the monolayer spectrum (black thick line) with the Fano fit (blue thin line) compared to a model neglecting electron-hole correlations (red dashed line).}
  \label{fig:data}
\end{figure*}

An example for the  absorption spectra acquired in this way  is given in Figure
Fig.~\ref{fig:data}A. For low photon energies, the spectra are
almost flat. The limiting values are close to integer multiples of
$\pi \alpha$  in accordance with
published results [3]. At higher energies, we find an asymmetric
peak at about 4.5 eV, as indicated by previous ellipsometric
measurements [11,12]. The Fano model, eq.~(\ref{eq:fano}), excellently describes our experimental data for both
monolayer as well as bilayer graphene over the entire recorded
spectral range. The single particle picture in a two-dimensional system results in a
symmetric fit function, eq.~(\ref{eq:singleparticle}). It   
utterly fails to describe the measured spectra (Fig.~\ref{fig:data}B, dashed line). In accordance with
ab initio calculations [13], the excitonic resonance carries all the
oscillator strength. The direct interband transition cannot be
identified.

The Fano model yields the excitonic resonance positions $E_r$ for monolayer and bilayer graphene (Table \ref{tab:fits}). The exciton binding energy is the energetic difference from the resonance to the van-Hove singularity calculated in the single-particle picture [13]. We find binding energies of about 400 meV and 250 meV for monolayer and bilayer graphene, respectively. Table \ref{tab:fits} summarizes the resulting fit parameters and characteristic deduced values. For common three-dimensional bulk metals, static screening by the electrons would prevent correlated electron-hole pairs within the Thomas-Fermi approximation. A reduced dimensionality, however, such as in one-dimensional metallic carbon nanotubes weakens the static screening so that excitonic correlations can occur [14]. The screening depends on the density of states near the Fermi energy. For the semi-metallic graphene, the density of states vanishes at the Dirac point and then grows linearly [15]. In bilayer graphene, however, the density of states is constant and nonzero near the Dirac point. For Fermi energies near the Dirac point, the screening ability of electrons in monolayer graphene is therefore hampered, and it is larger in bilayer graphene. As a result, the exciton binding energy is expected to be smaller when comparing bilayer with monolayer graphene in agreement with our measurements.

\begin{table}
\begin{tabular}{l|cccc|cc}
 dataset                 &  $E_r$ (eV)  & $\gamma$ (eV)  & $q$    &  C (\%)
& $A(0) /(\pi \alpha )$ & $E_{\rm{b}}$ (meV) \\
\hline monolayer$^\star$     &  4.73        &  1.30          &  -3.3  &  0.9
&  0.82                &  420\\
monolayer   &  4.78        &  1.58          &  -3.6  &  0.7
&  0.75                &  370 \\
bilayer$^\star$      &  4.70        &  1.63          &  -3.2  &  2.0
&  2.0                 &  270 \\
bilayer    &  4.73        &  1.39          &  -3.3  &  1.8
&  1.7                 &  240 \\
\end{tabular}
\caption{Summary of the fitting parameters and deduced values. The datasets marked with a star ($^\star$) are shown in Figure \ref{fig:data}.
$A(0)$ denotes the absorbance at zero energy as given by the model,
which has to be compared to integer multiples of $\pi \alpha$.
$E_{\rm{b}}$ is the exciton binding energy calculated as the
difference of $E_r$ to the saddle point.} \label{tab:fits}
\end{table}

It is intriguing that the Fano model describes the absorption
quantitatively correct over such a broad spectral range. It even
reproduces the absorption of $\pi \alpha$ in the low energy limit. According to the Fano model,
less than half of the absorption at low energies is due to the
constant continuum contribution associated with the Dirac cones. The
tail of the excitonic resonance still dominates at these low
energies. The success of the single resonance Fano model together
with the independence of the low energy absorption limit on the
slope of the Dirac cone [2,3] suggests that the detailed shape of
the band structure is less relevant, but rather the specific
topology of the monolayer and bilayer graphene band structure is
crucial.

We thank Th. Basch\'{e} for the loan of the Ultrafluar objective and S. Hein for visualizations.

\ \\

\textbf{References}
\setlength{\parindent}{0pt}

1. A. H. Castro Neto et al., \textit{Rev. Mod. Phys.} \textbf{81}, 109 (2009).

2. A. B. Kuzmenko et al., \textit{Phys. Rev. Lett.} \textbf{100}, 117401 (2008).

3. R. R. Nair et al., \textit{Science} \textbf{320}, 1308 (2008).

4. K. F. Mak et al., \textit{Phys. Rev. Lett}. \textbf{101}, 196405 (2008)

5. P. Y. Yu, M. Cardona, \textit{Fundamentals of Semiconductors} (Springer, Berlin, 2005).

6. J. C. Phillips, \textit{Excitons} in J. Tauc (ed.) \textit{The Optical Properties of Solids} (Academic Press, New York, 1966)

7. K. S. Novoselov et al., \textit{Science}
\textbf{306}, 666 (2004).

8. A. C. Ferrari et al., \textit{Phys. Rev. Lett.}
\textbf{97}, 187401 (2006).

9. J. Yan et al., \textit{Phys. Rev. Lett.} \textbf{98},
166802 (2007).

10. Z. H. Ni  et al., \textit{ACS Nano} \textbf{2}, 2301
(2008).

11. V. G. Kravets et al., \textit{Phys. Rev. B} \textbf{81}, 155413 (2010)

12. J. W. Weber et al., Appl. Phys. Lett. 97, 091904 (2010).

13. L. Yang et al. , \textit{Phys. Rev. Lett.} \textbf{103}, 186802 (2009).

14. F. Wang et al., \textit{Phys. Rev. Lett.} \textbf{99}, 227401 (2007).

15. T. Ando, \textit{J. Phys. Soc. Jpn.} \textbf{75}, 074716 (2006).

\end{document}